\documentclass[11pt,floatfix,amssymb,prx,aps,longbibliography,superscriptaddress,nopacs]{revtex4-2}

\usepackage{comment}
\newcommand{\om}{\left( \omega \right)}
\newcommand{\ta}{\tilde{a}\left(\omega\right)}

\newcommand{\tbx}{\tilde{b}_x\left(\omega\right)}
\newcommand{\tbxd}{\tilde{b^{\dagger}}_x\left(\omega\right)}

\newcommand{\tby}{\tilde{b}_y\left(\omega\right)}
\newcommand{\tbyd}{\tilde{b^{\dagger}}_y\left(\omega\right)}

\newcommand{\tZo}{\tilde{Z}\left(\omega\right)}
\newcommand{\txo}{\tilde{x}\left(\omega\right)}
\newcommand{\tyo}{\tilde{y}\left(\omega\right)}
\newcommand{\tNx}{\tilde{N}_x\left(\omega\right)}
\newcommand{\tNy}{\tilde{N}_y\left(\omega\right)}
\newcommand{\ccm}{\chi_\mathrm{c}^-\left(\omega\right)}
\newcommand{\cxm}{\chi_x^-\left(\omega\right)}
\newcommand{\cym}{\chi_y^-\left(\omega\right)}
\newcommand{\Ix}{\tilde{I}_x\left(\omega\right)}
\newcommand{\Iy}{\tilde{I}_y\left(\omega\right)}

\newcommand{\nm}{\bar{n}}
\newcommand{\OmegaX}{\Omega_{\scriptstyle{x}}}
\newcommand{\OmegaY}{\Omega_{\scriptstyle{y}}}
\newcommand{\OmegaZ}{\Omega_{\scriptstyle{z}}}
\newcommand{\Omegab}{\Omega_{\mathrm{b}}}
\newcommand{\Omegad}{\Omega_{\mathrm{d}}}
\newcommand{\Omegaeff}{\Omega_{\scriptstyle{\mathrm{eff}}}}
\newcommand{\GY}{\Gamma_{\scriptstyle{y}}}
\newcommand{\GX}{\Gamma_{\scriptstyle{x}}}

\newcommand{\omegaL}{\omega_{\scriptscriptstyle{\mathrm{L}}}}
\newcommand{\gX}{g_{\scriptstyle{x}}}
\newcommand{\gY}{g_{\scriptstyle{y}}}
\newcommand{\gb}{g_{\mathrm{b}}}
\newcommand{\bX}{\hat{b}_{\scriptstyle{x}}}
\newcommand{\bY}{\hat{b}_{\scriptstyle{y}}}
\newcommand{\bdX}{\hat{b}^{\dagger}_{\scriptstyle{x}}}
\newcommand{\bdY}{\hat{b}^{\dagger}_{\scriptstyle{y}}}
\newcommand{\ac}{\hat{a}_{\mathrm{c}}}

\newcommand{\Omegamj}{\Omega_{j}}
\newcommand{\xzpf}{x_{\mathrm{zpf}}}

\newcommand{\yzpf}{y_{\mathrm{zpf}}}
\newcommand{\Xc}{\mathrm{X}_{\scriptstyle{\mathrm{c}}}}
\newcommand{\Yc}{\mathrm{Y}_{\scriptstyle{\mathrm{c}}}}

\usepackage{amsmath}
\usepackage[dvips]{graphicx}
\usepackage{epsfig}
\usepackage{tabularx}
\usepackage{xcolor}
\begin{document}

\title{Spectral Analysis of Quantum Field Fluctuations in a Strongly Coupled Optomechanical System}

\author{A. Ranfagni}
\affiliation{Dipartimento di Fisica e Astronomia, Universit\`a di Firenze, via Sansone 1, I-50019 Sesto Fiorentino (FI), Italy}
\affiliation{European Laboratory for Non-Linear Spectroscopy (LENS), via Carrara 1, I-50019 Sesto Fiorentino (FI), Italy}
\affiliation{INFN, Sezione di Firenze, via Sansone 1, I-50019 Sesto Fiorentino (FI), Italy}

\author{F. Marino}
\affiliation{CNR-INO, largo Enrico Fermi 6, I-50125 Firenze, Italy}
\affiliation{INFN, Sezione di Firenze, via Sansone 1, I-50019 Sesto Fiorentino (FI), Italy}

\author{F. Marin}
\email[Electronic mail: ]{marin@fi.infn.it}
\affiliation{Dipartimento di Fisica e Astronomia, Universit\`a di Firenze, via Sansone 1, I-50019 Sesto Fiorentino (FI), Italy}
\affiliation{European Laboratory for Non-Linear Spectroscopy (LENS), via Carrara 1, I-50019 Sesto Fiorentino (FI), Italy}
\affiliation{INFN, Sezione di Firenze, via Sansone 1, I-50019 Sesto Fiorentino (FI), Italy}
\affiliation{CNR-INO, largo Enrico Fermi 6, I-50125 Firenze, Italy}

%\date{\today}

\begin{abstract}
With a levitodynamics experiment in the strong and coherent quantum optomechanical coupling regime, we demonstrate that the oscillator acts as a broadband quantum spectrum analyzer. The asymmetry between positive and negative frequency branches in the displacement spectrum traces out the spectral features of the quantum fluctuations in the cavity field, which are thus explored over a wide spectral range. Moreover, in our two-dimensional mechanical system the quantum back-action, generated by such vacuum fluctuations, is strongly suppressed in a narrow spectral region due to a destructive interference in the overall susceptibility.
\end{abstract}

\maketitle

A classical observable, described by a real function of time, always manifests a frequency-symmetric spectrum. This requirement does not hold for quantum observables, which can exhibit asymmetric spectra originated by non-commuting operators. The spectral asymmetry is therefore an important indicator of non-classicality, and a useful tool for investigating quantum properties. An important example of such quantum features is the vacuum noise in the photon number of the electromagnetic field in a driven cavity, which is indeed expected to present an asymmetric spectrum \cite{Clerk2010}. This characteristic can be shown by parametrically coupling the field to the position of a mechanical oscillator, in opto-mechanics experiments \cite{Aspelmeyer2014,Bowen2015}. In the weak coupling regime ($g < \kappa/4$, where $g$ is the optomechanical coupling rate and $\kappa$ is the  optical decay rate), the interaction between electromagnetic field and mechanical motion can be understood in the framework of a Raman scattering picture, with an anti-Stokes (and a Stokes) process transferring energy and quanta from the oscillator to the cavity field (and vice-versa) \cite{Marquardt2007,Wilson-Rae2007}. As a result of this process, two motional sideband peaks are imprinted in the spectrum of the cavity field, tracing out the displacement spectrum of the oscillator. Due to the asymmetry in the field vacuum fluctuations, the two sidebands are unbalanced, with the anti-Stokes (Stokes) Lorentzian peak amplitude proportional to $\nm$ ($\nm + 1$). Here, $\nm$ is the resulting phononic occupancy of the mechanical oscillator \cite{Safavi-Naeini2012,Khalili2012} which, for suitable detuning $\Delta$ between the driving field and the cavity mode, is deeply cooled thanks to the different probabilities of the two processes. The sideband asymmetry can be appreciated in a heterodyne detection of the field exiting the cavity \cite{Purdy2015,Underwood2015}. In this regime, the oscillator motion just couples to the cavity field in a narrow spectral region around the mechanical resonance. The amplitude imbalance between the anti-Stokes and Stokes peaks is a signature of both the spectral asymmetry of the back-action provided by the field vacuum fluctuations, and the non-classical motion of the oscillator approaching its ground state \cite{Weinstein2014,Borkje2016}. 

We argue that the analysis of the frequency asymmetry in the full spectrum, extended beyond a mere quantification of the amplitude ratio between similar Lorentzian peaks, is a potentially powerful tool for exploring the quantum behavior of optomechanical systems. 
Already in the weak coupling regime, different shapes of the two sidebands have been observed in a parametric squeezing experiment \cite{Chowdhury2020,Vezio2020}. More generally, the full spectral analysis is particularly efficient outside the regime of single-mode and weak optomechanical coupling, where the scattering picture is no more sufficient to explain the displacement spectrum and, consequently, the spectrum of the output field.

In this work we exploit the quantum-coherent strong coupling regime \cite{Teufel2011,Verhagen2012,Wollman2015,Pirkka2015,Ranfagni2021} and show that the oscillator fully unfolds its potentialities as quantum spectrometer \cite{Clerk2010}. The difference between the positive and negative frequency branches in the displacement spectrum is traced back to the characteristics of the quantum fluctuations in the cavity field, in particular their spectral shape dictated by the cavity filtering. Moreover, in a two-dimensional mechanical system we demonstrate that the quantum back-action, generated by the vacuum field fluctuations, is cancelled in a narrow spectral region, due to the structure of the overall susceptibility characterized by modal interferences. Such cancellation does not occur for the thermal noise, since different modes couple to uncorrelated baths \cite{Caniard2007}. As a consequence, the spectral asymmetry, which singles out the quantum component of the back-action, exhibits a sharp characteristic dip which provides the clearest experimental signature of the field vacuum fluctuations. Furthermore, for strong enough two-dimensional cooling, the Stokes sideband, dominated by the quantum backaction, presents a peculiar hole.   

\begin{figure}[!htb]
    \centering
    \includegraphics[scale=1]{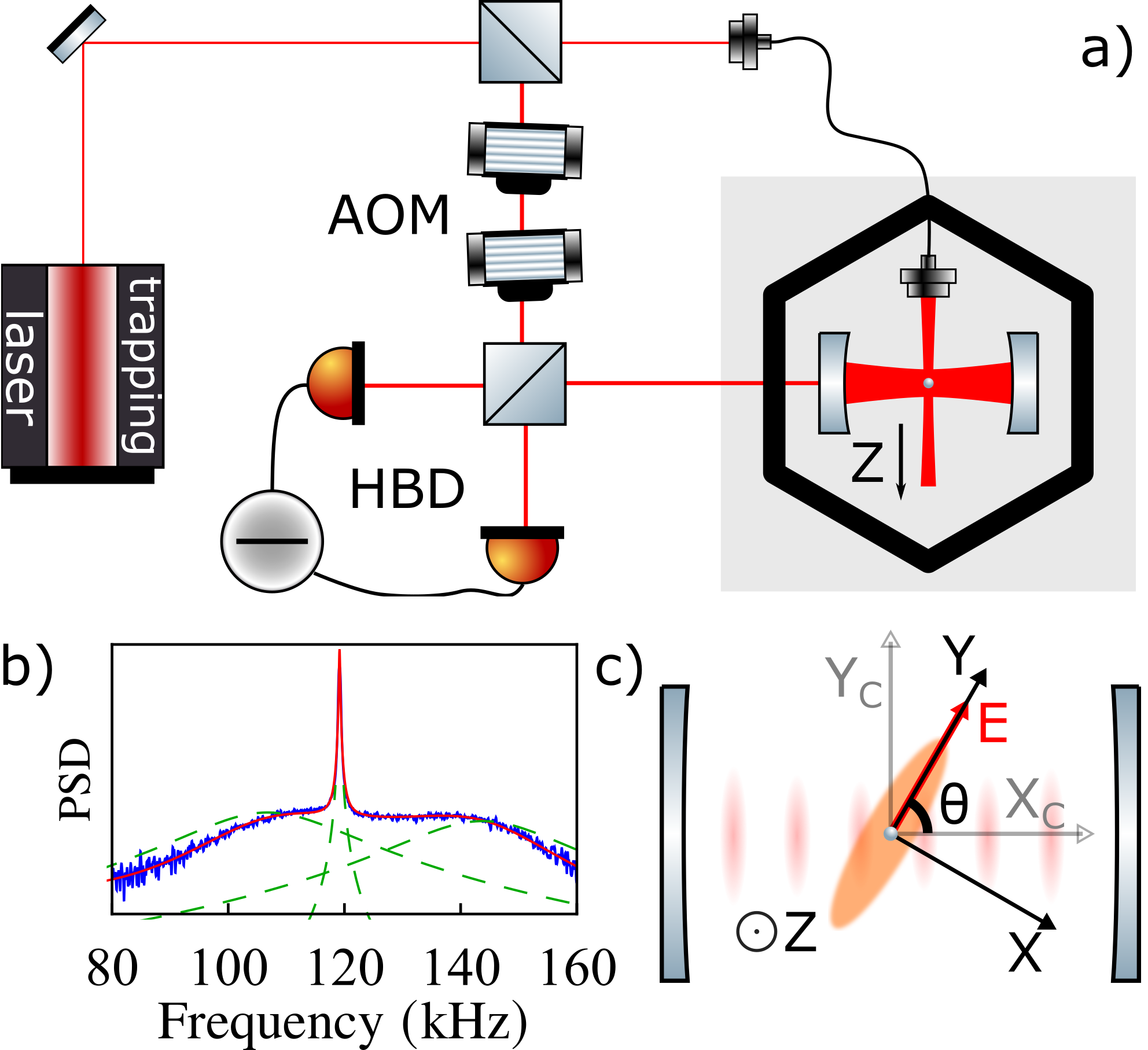}
    \caption[setup]{a) Simplified scheme of the experimental setup. AOM: acousto-optic modulator, HBD: balanced heterodyne detection. b) Power spectral density (PSD) of the heterodyne signal (anti-Stokes sideband), normalized to shot noise. A narrow peak due to the coupling between bright and dark modes is visible between two broad polaritonic resonances \cite{Ranfagni2021}. To emphasize this structure, we show with dashed lines three Lorentzian shapes with centers and widths determined by the system eigenvalues, and amplitudes adapted to fit the experimental spectrum. The red solid line is the fit with the full model of Eqs. (\ref{eq_Sout}) and (\ref{eq:psd_GBM}). The parameters are the same as in Fig. \ref{Fig_asym}, with a detuning $\Delta/2\pi = -130\,$kHz and a polarization angle $\theta = 81^{\circ}$. c) Schematic of the plane orthogonal to the tweezer axis. In the tweezer frame, Y denotes the tweezer polarization axis and X its orthogonal direction. In the cavity frame, $\Xc$ denotes the cavity axis, and $\Yc$ its orthogonal direction.}
    \label{fig:setup}
\end{figure}

Our experimental system [Fig. \ref{fig:setup}(a)] is based on a silica nanosphere (diameter 125 nm) levitated in the optical potential created by a strongly focused laser beam, generated by a Nd:YAG source (optical tweezer \cite{Ashkin1970}) inside a vacuum chamber \cite{Millen2020,Gonzales-Ballestero2021}. Labelling with Z the tweezer axis and Y the axis of the linear polarization [see Fig. \ref{fig:setup}(c) for a scheme], the oscillation frequencies are $(\OmegaX, \OmegaY, \OmegaZ)/2\pi = (131,120,30)\,$kHz. The loading procedure is described in Ref. \cite{Calamai2021}. Due to its much lower frequency, the motion along Z is completely decoupled from the one on the orthogonal plane, and we will neglect it in this article. On the other hand, the motion on the XY plane is described by a fully two-dimensional system, as $(\OmegaX - \OmegaY)$ is comparable to the optomechanical shifts and broadenings. The nanosphere is accurately positioned on the axis of a Fabry-Perot cavity (linewidth $\kappa/2\pi = 57\,$kHz), which is almost orthogonal to the tweezer axis. The light of the tweezer has an accurately controlled detuning $\Delta$ from a cavity resonance, and the nanosphere is placed in correspondence of a node of the corresponding cavity standing wave. The motion along the cavity axis is coupled to the cavity mode by coherent scattering \cite{Vuletic2000,Windey2019,Delic2019A}. This technique provides a large optomechanical coupling, which recently allowed cooling the nanosphere oscillations down to phononic occupancy below unity \cite{Delic2020,Ranfagni2022,Piotrowski2022}. The tweezer light scattered into the cavity mode and transmitted by the end mirror is analyzed in a balanced heterodyne detection. Further details on the experimental setup are given in Refs. \cite{Ranfagni2021,Ranfagni2022}.

The linearized evolution equations for the motion in the plane orthogonal to the tweezer axis, expressed in the frame rotating at the laser frequency $\omegaL$, can be written as 
\begin{equation}
\dot{\hat{a}}_{\mathrm{c}}=\bigg(  i\Delta-\frac{\kappa}{2}\bigg)\ac+i\gX(\hat{b}_x+\hat{b}_x^\dag)+i\gY(\hat{b}_y+\hat{b}_y^\dag)+\sqrt{\kappa}\,\hat{a}_{\mathrm{in}}
\label{dta}
\end{equation}
\begin{equation}
\dot{\hat{b}}_j=\left(-i\Omegamj-\frac{\gamma_j}{2}\right)\hat{b}_j+ig_j(\ac+\ac^{\dag})+\sqrt{\Gamma_j}\,\hat{b}_{\mathrm{n},j} 
\label{dtb}
\end{equation}
where the operators $\ac$ and $\hat{b}_j$ describe respectively the intracavity field and the two mechanical modes ($j=x, y$), $\gamma_j$ are the gas damping rates, and $g_j$ are the optomechanical coupling rates. The ladder operators $\hat{b}_j$ are linked to the operators describing the displacements ($\hat{x}$, $\hat{y}$) by the relations
$\hat{x} = \xzpf (\bX+\bdX)$ and $\hat{y} = \yzpf (\bY+\bdY)$,
where $\,\xzpf = \sqrt{\hbar/2 m \OmegaX}$ and $\,\yzpf = \sqrt{\hbar/2 m \OmegaY}$ are the zero-point position fluctuations of the free oscillators, and $m$ is the mass of the nanosphere.  
The input noise operators are characterized by the correlation functions
$\langle\hat{a}_{\mathrm{in}}(t)\hat{a}_{\mathrm{in}}^{\dag}(t')\rangle  =  \delta(t-t')$, $\langle\hat{a}_{\mathrm{in}}^{\dag}(t)\hat{a}_{\mathrm{in}}(t')\rangle  =  0$, $\langle\hat{b}_{\mathrm{n},j}^{\dag}(t)\hat{b}_{\mathrm{n},j}(t')\rangle = \langle\hat{b}_{\mathrm{n},j}(t)\hat{b}_{\mathrm{n},j}^{\dag}(t')\rangle  =  \delta(t-t')$.
The total decoherence rates $\Gamma_j$ are due to collisions with the background gas molecules \cite{Beresnev1990},  
and to the shot noise in the dipole scattering \cite{Seberson2020}. For both processes we are using a classical description, justified by the oscillation frequencies and the operation at room temperature. 

We note that the optomechanical coupling rates can be written as $\gX = g_{\mathrm{max}}\,\sin^2 \theta$ 
and $\gY = g_{\mathrm{max}}\,\sqrt{\frac{\OmegaX}{\OmegaY}}\sin \theta \cos \theta$ \cite{Delic2019A}, 
where $\theta$ is the angle between the cavity axis $\Xc$ and the tweezer polarization axis [see Fig. \ref{fig:setup}(c)]. The optomechanical coupling with the motion perpendicular to the cavity axis is null and indeed in our model this coupling is proportional to  $\propto \left( \sin \theta\,\gY/\yzpf \,-\,\cos \theta \, \gX/\xzpf \right) \,=\, 0$. Whenever the effective width of the mechanical resonances, increased by the optomechanical coupling, is comparable to or larger than the frequency splitting $(\OmegaX - \OmegaY)$, it is useful to describe the planar motion using the axis $\Xc$ and $\Yc$ (cavity frame). The motions along these directions define respectively the so-called geometrical bright and dark modes \cite{Toros2021}, with frequencies given respectively by
\begin{eqnarray}
\Omegab^2 = \sin^2 \theta \,\OmegaX^2 + \cos^2 \theta \,\OmegaY^2  \\
\Omegad^2 = \cos^2 \theta \,\OmegaX^2 + \sin^2 \theta \,\OmegaY^2 
\end{eqnarray}
and optomechanical coupling to the bright mode
$\gb=\sqrt{(\gX^2\OmegaX+\gY^2 \OmegaY)/\Omegab} = g_{\mathrm{max}} \sin \theta \sqrt{\OmegaX/\Omegab}$.
In the strong coupling regime, the spectrum of the bright mode exhibits two broad peaks corresponding to the polaritonic resonances [see Fig.  \ref{fig:setup}(b)]. Since the $x$ and $y$ frequencies are not degenerate, the bright and dark directions do not correspond to eigenstates of the total Hamiltonian, therefore the spectrum of the bright mode also shows a third peak, corresponding to the third eigenstate of the three-dimensional optomechanical system. In the cavity frame it can be ascribed to the coupling between bright and dark modes and, in loose terms, it is called dark mode peak.

The total cavity output field is given by the input-output relation $\hat{a} = \sqrt{\kappa} \hat{a}_{\mathrm{c}}-\hat{a}_{\mathrm{in}}$. The heterodyne spectrum normalized to shot noise can be written as
\begin{equation}
S_{\mathrm{out}}(\Omega_{\mathrm{LO}}+\omega) 
= 1 + \eta \,\gb^2 \,\kappa \,\left|\chi_\mathrm{c}\left(\omega\right)\right|^2 S_{x_\mathrm{b} x_\mathrm{b}}\om
\label{eq_Sout}
\end{equation}
where 
$\Omega_{\mathrm{LO}}$ is the angular frequency of the local oscillator, and $\eta$ is the overall detection efficiency. The displacement spectrum $S_{x_\mathrm{b} x_\mathrm{b}}$ of the bright mode appears, filtered by the optical susceptibility $\chi_\mathrm{c}=\left[-i(\Delta+\omega)+ (\kappa/2)\right]^{-1}$. To recover the main system features from the measured output spectrum, we thus calculate a corrected asymmetry defined for $\omega > 0$ as
\begin{equation}
A\om\,=\,\frac{S_{\mathrm{out}} (\Omega_{\mathrm{LO}}-\omega)\,-\,1}{S_{\mathrm{out}} (\Omega_{\mathrm{LO}}+\omega)\,-\,1}\,\,\frac{(\omega-\Delta)^2+(\kappa/2)^2}{(\omega+\Delta)^2+(\kappa/2)^2}  \, .
\label{Eq_asym}
\end{equation}
According to Eq. (\ref{eq_Sout}), it provides the imbalance between the negative and frequency branches of the displacement spectrum, i.e., $A\om \equiv S_{x_\mathrm{b} x_\mathrm{b}}\left(-\omega\right)/S_{x_\mathrm{b} x_\mathrm{b}}\om$.

\begin{figure*}[!htb]
    \centering
    \includegraphics[scale=1]{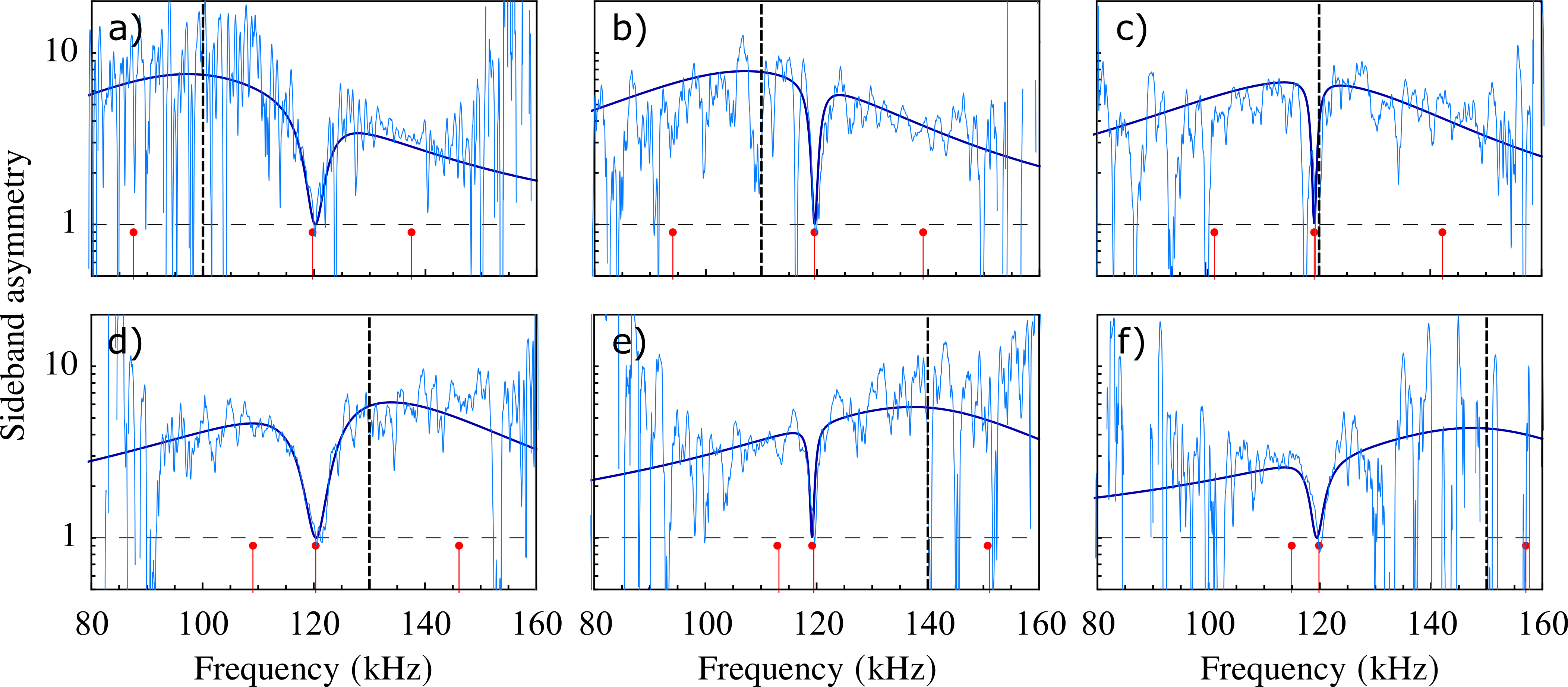}
    \caption[Sideband asymmetry in the strong coupling regime]{Spectral asymmetry $A(\omega)$ for a tweezer light detuning ranging from $-\Delta / 2 \pi = 100 \,\mathrm{kHz} $ (panel a) to $-\Delta / 2 \pi = 150 \,\mathrm{kHz} $ (panel f) in steps of 10 kHz, at a background pressure of $4 \times 10^{-6}\,\mathrm{Pa}$. The detuning is indicated by the vertical dashed lines. The dark blue solid lines are produced by the theoretical model, fitted to the experimental data. The fits give mean values of the maximum optomechanical coupling rate $g_{\mathrm{max}} / 2 \pi = 23.5 \ \mathrm{kHz}$, of the  decoherence rates $\GX/ 2\pi = 6.2 \ \mathrm{kHz}$ and $\GY/ 2 \pi = 5.6 \ \mathrm{kHz}$, and polarization angles $\theta = (71, 81, 84, 67, 84, 71)^{\circ}$. The red dots display the eigenfrequencies of the system extracted from the drift matrix.
    }
    \label{Fig_asym}
\end{figure*} 

From the system equations, we can obtain the stationary power spectrum $S_{x_\mathrm{b} x_\mathrm{b}}$, which reads \cite{SM}:
\begin{multline}
S_{x_\mathrm{b} x_\mathrm{b}}\om=\frac{1}{\gb^2}\frac{1}{\left|1+\chi_\mathrm{c}^{-}
\left(\gX^2 \chi_x^{-}+\gY^2 \chi_y^{-}\right)\right|^2}  \\
\Bigr[
\gX^2 \GX \left(
\left|\chi_x\om\right|^2+\left|\chi_x\left(-\omega\right)\right|^2\right)
%+ \\
+\gY^2 \GY \left(
\left|\chi_y\om\right|^2+\left|\chi_y\left(-\omega\right)\right|^2\right)
+\\
+
\left|\gX^2 \chi_x^{-}+\gY^2 \chi_y^{-}\right|^2
\kappa \left|\chi_\mathrm{c}\left(-\omega\right)\right|^2\Bigr],
\label{eq:psd_GBM}
\end{multline}
where $\chi_\mathrm{c}^{-}=\chi_\mathrm{c}\om-\chi_\mathrm{c}^*\left(-\omega\right)$, $\chi_{j}^{-}  =\chi_{j}\om-\chi_{j}^*\left(-\omega\right)$, and $\chi_{j}\om=\left[i(\Omegamj-\omega)+ (\gamma_{j}/2)\right]^{-1}$ are the mechanical susceptibilities.
In Eq. (\ref{eq:psd_GBM}) we can distinguish the contribution
 of the spectrally flat
noise forces quantified by the decoherence rates $\Gamma_j$, from that generated by
the optical vacuum noise, filtered by the cavity
and therefore proportional to $\kappa \left|\chi_\mathrm{c}\left(-\omega\right)\right|^2 $.

The nanosphere is levitated in high 
vacuum, where its dynamics is 
dominated by the strong coupling with the cavity field  and 
markedly non-classical effects can be unveiled. The quantum
features are highlighted in the
spectral asymmetry
reported in  Fig. \ref{Fig_asym}
for different values of the detuning.
The ratio between the sidebands strongly departs from the classical unit value on a broad spectral region.
The maximal
asymmetry occurs for frequencies 
around $\omega \sim -\Delta$, where it
ranges between $\sim 5$ and $\sim 8$, denoting a strong non-classical behavior.

For the interpretation of these spectral features it is instructive
to start from the one-dimensional limit of Eq. (\ref{eq:psd_GBM}), obtained for $\theta = \pi/2$ and therefore $\gY = 0$, which gives the spectrum:

\begin{equation}
\label{Sxx_simpl}
S_{x x} \om \simeq  \left|\chi_\mathrm{m}^{\mathrm{eff}}\left(\omega \right)\right|^2  
\left[
 \Gamma 
+
\kappa g^2 \left|
\chi_\mathrm{c}\left(
-\omega
\right)
\right|^2 
\right]
\end{equation}
where, for the seek of clarity, we have set $x_\mathrm{b}\to x$,
$\GX \to \Gamma$, $\gX \to g$, and
where $\chi_\mathrm{m}^{\mathrm{eff}}$  is the effective mechanical susceptibility, with $\left|\chi_\mathrm{m}^{\mathrm{eff}}\left(-\omega \right)\right|^2 = \left|\chi_\mathrm{m}^{\mathrm{eff}}\left(\omega \right)\right|^2$.
The spectral asymmetry is determined by the term proportional to $\left|
\chi_\mathrm{c}\left(
-\omega
\right)
\right|^2$, and it is particularly relevant at detuning $-\Delta \simeq \OmegaX$, for large quantum cooperativity $C_{\scriptscriptstyle{\mathrm{Q}}}=\frac{4 g^2}{\kappa \Gamma} \gg 1$. 
This is also the requirement for achieving ground state cooling of the mechanical oscillator. In the weak coupling regime, $\chi_\mathrm{m}^{\mathrm{eff}}\left(\omega \right)$ 
gives a couple of Lorentzian peaks much narrower than $\kappa$, centered at frequencies $\pm\Omegaeff$ 
shifted with respect to $\pm \OmegaX$ due to the optical spring effect. In Eq. (\ref{Sxx_simpl}) we can therefore approximate 
$\chi_\mathrm{c}\left(
-\omega
\right)$  with its values at $\mp \Omegaeff$, obtaining different scaling factors for the Lorentzian Stokes and anti-Stokes peaks. 

In the strong coupling regime,
the effective mechanical susceptibility gets broader
and is composed of the two polaritonic peaks of
width $\sim \kappa/2$ \cite{Teufel2011,Ranfagni2021}.
The mechanical transfer function probes
the noise bath on a bandwidth of the order
of $\sim 2 g + \kappa/2 > \kappa$ around
the mechanical bare frequency on both the 
positive and negative frequency branches.
In this case, the spectral asymmetry 
can be experimentally investigated over a wide range and reads:
\begin{equation}
\frac{S_{x x}\left(-\omega\right)}{S_{x x}\left(\omega\right)}=\frac{S_{FF}\left(-\omega\right)}{S_{FF}\left(\omega\right)}=
\frac{\Gamma 
+
\kappa g^2 \left|
\chi_\mathrm{c}\left(
\omega
\right)
\right|^2}{\Gamma 
+
\kappa g^2 \left|
\chi_\mathrm{c}\left(
-\omega
\right)
\right|^2} \, .
\label{eq:asy:cav}
\end{equation} 
The oscillator acts as a broadband quantum spectrum 
analyzer of the external force, expressed in Eq. (\ref{eq:asy:cav}) in terms
of the power spectral density $S_{FF}\om$ \cite{Clerk2010}. As shown in Fig. \ref{Fig_asym}, thanks to the strong coupling and the high quantum cooperativity we can indeed appreciate the full shape of $A(\omega)$, that is dictated by the optical susceptibility and, as already observed and made clear by the right hand side of Eq. (\ref{eq:asy:cav}), reaches its maximum for $\omega \simeq -\Delta$. 

Returning to the two-dimensional system, a particular feature of the asymmetries, well visible in Fig. \ref{Fig_asym}, is a dip occuring at $\sim 119\,$kHz, regardless of the detuning. Here the two
sidebands have almost equal amplitudes and  their ratio  falls very close to unity. This peculiarity
derives from the presence of the two mechanical 
susceptibilities in the full system model.
The pre-factor $\gY^2 \chi_y^{-}+\gX^2 \chi_x^{-}$ in the last term of Eq. (\ref{eq:psd_GBM}) describes the superposition
of the linear responses of the two bare oscillators 
to the common mode optical radiation pressure
force. The two susceptibilities are weighted 
with their respective coupling rates.
In the spectral region between the mechanical
frequencies, the two oscillators react to the fluctuating optical force with opposite phase, leading to
a destructive interference effect. Remarkably, here
the quantum back action is inhibited from 
entering the mechanical system. With high mechanical quality oscillators, the asymmetry falls to $\sim 1$ because of the mechanical coupling to 
the flat, classical noise forces that here dominate.
We note that the antiresonance frequency matches the 
bare geometrical dark mode frequency $\Omegad$. The expected width of the dip is weakly dependent on the detuning, but it is sensitive to the balance between the two optomechanical coupling rates, achieving its maximum for $\gX \simeq \gY$. In the spectra shown in Fig. \ref{Fig_asym} the polarization angle $\theta$ fluctuates due to slow drifts in the fiber, and this explains the observed (and well fitted by the model) variations of the dip width. 

The interference dip provides the clearest experimental signature of the quantum behavior of the optomechanical system: since it appears on the spectral asymmetry we can rule out classical effects, and it is very weakly sensitive to uncertainties in the calibration and correction of the spectra for cavity filtering  (the depth of the minimum can indeed be used as a check for their accuracy).   

\begin{figure}[!htb]
    \centering
    \includegraphics[scale=1]{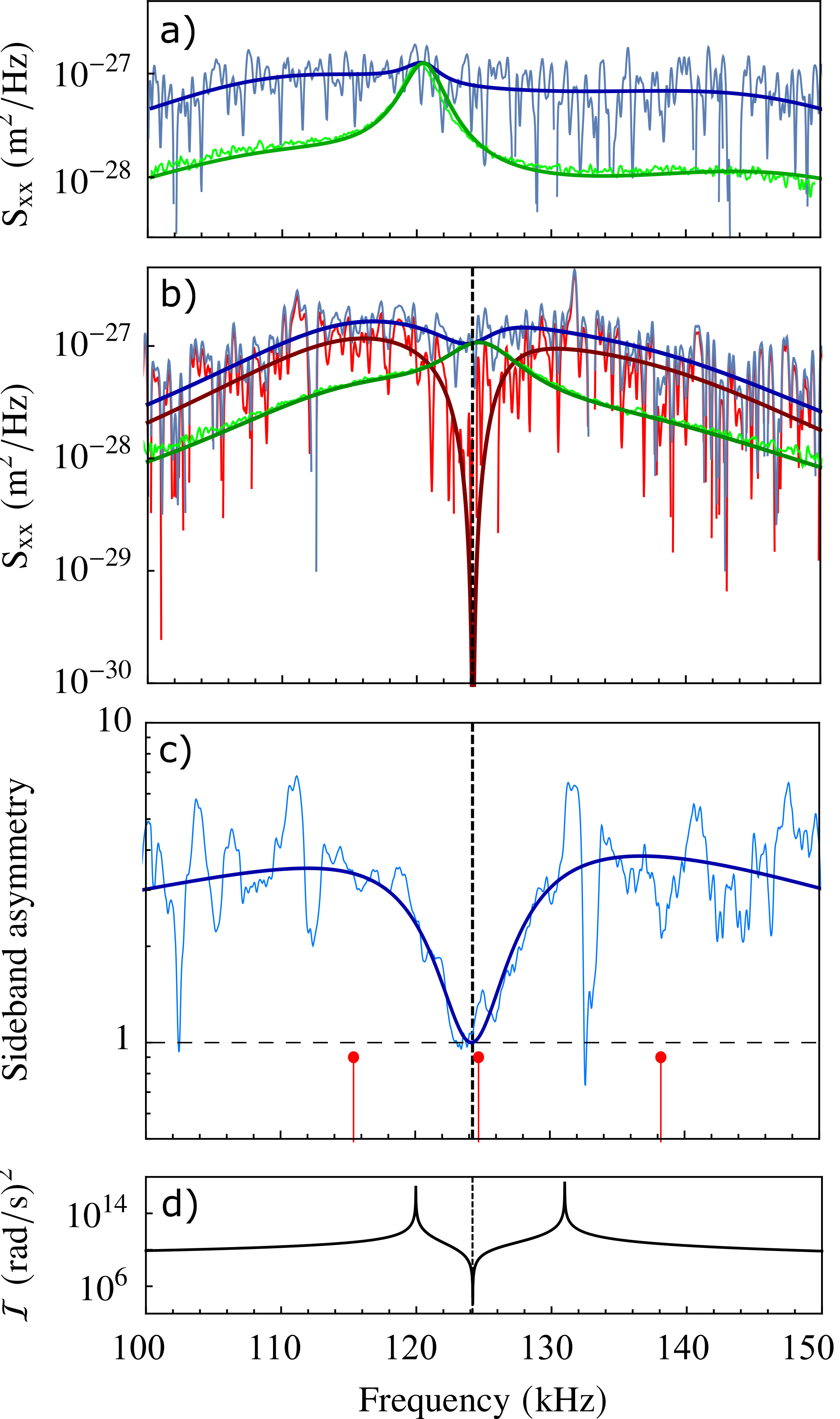}
    \caption{a): Stokes (light blue) and anti-Stokes (light green) sidebands, corrected
    for the cavity filtering, with the respective fitting functions (dark solid lines). The data are the same as in Fig. \ref{Fig_asym}(d). In the following panels, the polarization angle is decreased to $\theta = 52^{\circ}$.  b): The quantum back action contribution to the Stokes sideband is 
    plotted in light red, with the theoretical curve
    superimposed (dark red solid line).
    At the resonance  frequency of the geometrical dark mode (vertical dashed line) the quantum noise term is dynamically cancelled and the
    two spectral branches, here both driven only by the classical flat noise, overlap.
    On broad frequency regions at the sides of the antiresonance,
    the Stokes sideband is mostly
    determined  by the quantum fluctuations. The system parameters extracted from the fit are  $\gX/ 2\pi = 14 \ \mathrm{kHz}$, $\gY/ 2 \pi = 11 \ \mathrm{kHz}$, $\GX/ 2\pi = 5.8 \ \mathrm{kHz}$ and $\GY/ 2 \pi = 5.6\ \mathrm{kHz}$.  c): Asymmetry between the Stokes and anti-Stokes
    sidebands, for the spectra reported in panel (b). The system eigenfrequencies, extracted from the 
    drift matrix, are displayed with red dots.
    d): Spectral dependence of the interference term $\mathcal{I}=\left|\gX^2 \chi_x^{-}+\gY^2 \chi_y^{-}\right|^2$, 
    which exhibits a sharp dip at the frequency of the dark mode.}
    \label{Fig_2dcooling}
\end{figure}

We further analyze the broadband, cavity mediated back action noise 
in the experimental configuration
with optimal detuning $\Delta/2\pi = - 130\,$kHz and
 $\gX \simeq \gY$ (i.e., $\theta \simeq \pi/4$).
The Stokes and anti-Stokes sidebands,
converted to displacement spectrum according to Eq. (\ref{eq_Sout}),
are shown in Fig. \ref{Fig_2dcooling}(b) with the corresponding fitting functions. In the previous configuration, Stokes and anti-Stokes sidebands had similar shapes [see Fig. \ref{Fig_2dcooling}(a)], and the dip in the asymmetry was produced by a different balance between the amplitude of the dark mode peak (which, dominated by thermal noise, is similar in the two sidebands) and those of the polaritonic peaks (which, dominated by quantum noise, are much larger in the Stokes sideband). Here, on the contrary,
the two spectra have markedly different
patterns, with the Stokes one showing
a dip in the region of the dark mode resonance. This further, peculiar effect is obtained thanks to the strong two-dimensional cooling \cite{SM}. 

The quantum noise cancellation is
here effective on a broad frequency range, allowing a deeper analysis. 
While the classical flat noise gives 
the same spectral contribution on the two sidebands,
the back action noise
is highly suppressed
in the anti-Stokes branch,
because of the strong cavity
filtering, with $\left|\chi_\mathrm{c}\left(\Delta\right)/\chi_\mathrm{c}\left(-\Delta\right)\right|^2\sim 0.01$. The
difference between the negative and positive branches of the heterodyne spectrum, corrected for the cavity filtering, provides
the contribution to the motional
spectrum originated by the cavity quantum back
action noise. We plot it with a light red line in Fig. \ref{Fig_2dcooling}(b) and 
compare it with the model (solid dark line).
We see that the quantum force
exceeds the classical noise
on a wide spectral range
around the polaritonic resonance frequencies, displayed by the two external red dots
in Fig. \ref{Fig_2dcooling}(c). On the other hand,
close to the dark mode eigenfrequency
[central red dot in Fig. \ref{Fig_2dcooling}(c)],
the optical quantum noise is highly
suppressed and just the residual classical
noise drives the motion.
The cancellation of the back action noise 
had already been
explored for optomechanical systems with classical laser noise \cite{Caniard2007}, but we
remark that here we can show
the inhibition of the quantum back action, generated by vacuum fluctuations, 
thanks to an oscillator very close to its ground state.

The solid curve in Fig. \ref{Fig_2dcooling}(d) reports the
frequency dependence of the interference term $\mathcal{I}=\left|\gX^2 \chi_x^{-}+\gY^2 \chi_y^{-}\right|^2$.
The antiresonance occurs at the
bare geometrical dark mode frequency,
which differs from 
the corresponding eigenfrequency because of the non-degeneracy of the $x$ and $y$ modes \cite{Toros2021}.
Starting from $\sim 1$ at $\Omegad$,
the asymmetry grows and reaches its maximum 
around the polaritonic eigenfrequencies [Fig.\ref{Fig_2dcooling}(c)] .

In conclusion, we demonstrate that, in the quantum-coherent strong coupling regime, a mechanical oscillator acts as a spectrum analyzer of the intracavity field quantum fluctuations over a wide frequency range. We also show that, owing to the two-dimensional motion, the quantum back-action generated by vacuum field fluctuations is cancelled in a narrow spectral region due to destructive interference between the two bare mechanical resonances, with meaningful implications in quantum-limited sensing of weak forces \cite{Ranjit2016}.

\bibliography{database}

\newpage

\section*{Supplemental material}

\subsection*{Derivation of the analytical solution}
The Langevin model described in the main text can be solved in the Fourier space, for the field and motion quadratures defined as 
\begin{gather}
\tZo = \ta + \tilde{a}^{\dagger}\left(\omega\right) \\
\txo = \tbx + \tbxd \\
\tyo = \tby + \tbyd 
\end{gather}
where the tilde denotes Fourier transformed (FT) operators $\tilde{O}\left(\omega\right)=\mathrm{FT}\left[\hat{O}\left(t\right)\right]$ and
$\tilde{O}^{\dagger}\left(\omega\right)=\mathrm{FT}\left[\hat{O}^{\dagger}\left(t\right)\right]$.
The model can be written as
\begin{gather}
\tZo=i \ccm \left[g_x \txo+g_y \tyo  \right]+\tilde{Z}_\mathrm{in} \left(\omega\right)\\
\txo = i  \gX \cxm \tZo + \tNx \\
\tyo = i  \gY \cym \tZo + \tNy 
\end{gather}
where we have defined the susceptibilities
\begin{gather}
\chi_x\left(\omega\right)=\frac{1}{i
\left(\Omega_{x}-\omega\right)+\gamma_x/2}\\
\chi_y\left(\omega\right)=\frac{1}{i
\left(\Omega_{y}-\omega\right)+\gamma_y/2}\\
\chi_\mathrm{c}\left(\omega\right)=\frac{1}{-i
\left(\Delta+\omega\right)+\kappa/2}
\end{gather}
and the compact forms
\begin{gather}
\ccm=\chi_\mathrm{c}\left(\omega\right)-
\chi_\mathrm{c}^{*}\left(-\omega\right)\\
\cxm=\chi_x\left(\omega\right)-
\chi_x^{*}\left(-\omega\right)\\
\cym=\chi_y\left(\omega\right)-
\chi_y^{*}\left(-\omega\right) \, .
\end{gather}
The noise input terms can be written as
\begin{gather}
\tilde{Z}_\mathrm{in} \left(\omega\right) = \sqrt{\kappa}\left[\chi_\mathrm{c} \om \tilde{a}_{\mathrm{in}}\om +
\chi_\mathrm{c}^{*} \left(- \omega\right) \tilde{a}^{\dagger}_{\mathrm{in}}\left(  \omega \right)  \right]\\
\tNx = \sqrt{\GX}\left[\chi_x \om \tilde{b}_{\mathrm{in},x}\om +
\chi_x^{*} \left(- \omega\right) \tilde{b}^{\dagger}_{\mathrm{in},x}\left(  \omega \right)  \right] \\
\tNy = \sqrt{\GY}\left[\chi_y \om \tilde{b}_{\mathrm{in},y}\om +
\chi_y^{*} \left(- \omega\right) \tilde{b}^{\dagger}_{\mathrm{in},y}\left(  \omega \right)  \right]  \, .
\end{gather}

Solving the system, we obtain the position along the two axes in terms of the noise input terms:
\begin{gather}
\label{eq:xy1}
\txo = \frac{1}{\Ix + \Iy -1} \left[ i \gX \cxm \tilde{Z}_\mathrm{in} \left(\omega\right) + \Iy \tNx -
\left(\Ix-1\right) \frac{\gY}{\gX}\tNy \right] \\
\label{eq:xy2}
\tyo = \frac{1}{\Ix + \Iy -1} \left[ i \gY \cym \tilde{Z}_\mathrm{in} \left(\omega\right) + \Ix \tNy -
\left(\Iy-1\right) \frac{\gX}{\gY}\tNx \right]
\end{gather}
with
\begin{gather}
\Ix = 1+ \gX^2 \cxm \ccm\\
\Iy = 1+ \gY^2 \cym \ccm  \,  .
\end{gather}

The geometrical bright mode, corresponding to the motion along the cavity axis, 
is defined in terms of the parameters used in the model by the coordinate $x_\mathrm{b} = \frac{1}{g_\mathrm{b}}\left(\gX x+ \gY y\right)$, where
\begin{gather}
g_\mathrm{b}=\sqrt{\frac{\gX^2\Omega_x+\gY ^2 \Omega_y}{\Omega_{\mathrm{b}}}}\\
\Omega_{\mathrm{b}}=\sqrt{\frac{\gX^2 \Omega_{x}^3+\gY^2 \Omega_{y}^3}{\gX^2\Omega_{x}+\gY^2 \Omega_{y}}}  \, .
\end{gather}
Plugging in the expressions (\ref{eq:xy1}-\ref{eq:xy2}) we derive
\begin{equation}
\tilde{x}_\mathrm{b} \left( \omega \right)=\frac{1}{g_\mathrm{b}\left[\Ix + \Iy -1\right]}\Bigl\{i\left[\gX^2\cxm +\gY^2 \cym \right] \tilde{Z}_\mathrm{in}  \left(\omega\right) +\gX \tNx + \gY \tNy \Bigr\}
\end{equation}
from which we calculate the displacement noise spectrum
\begin{multline}
S_{x_\mathrm{b} x_\mathrm{b}}\om= \frac{1}{g_\mathrm{b}^2}\frac{1}{\left|1+\chi_\mathrm{c}^{-}
\left(\gX^2 \chi_x^{-}+\gY^2 \chi_y^{-}\right)\right|^2}\cdot  
\Bigr[
\gX^2 \GX \left(
\left|\chi_x\om\right|^2+\left|\chi_x\left(-\omega\right)\right|^2\right)+\\
+\gY^2 \GY \left(
\left|\chi_y\om\right|^2+\left|\chi_y\left(-\omega\right)\right|^2\right)
+
\left|\gX^2 \chi_x^{-}+\gY^2 \chi_y^{-}\right|^2
\kappa \left|\chi_\mathrm{c}\left(-\omega\right)\right|^2\Bigr]
\label{eq:psd_GBM}
\end{multline}
used in the main text.
\newpage

\subsection*{Spectrum of the Stokes sideband}

\begin{figure*}[!htb]
    \centering
    \includegraphics[scale=1.6]{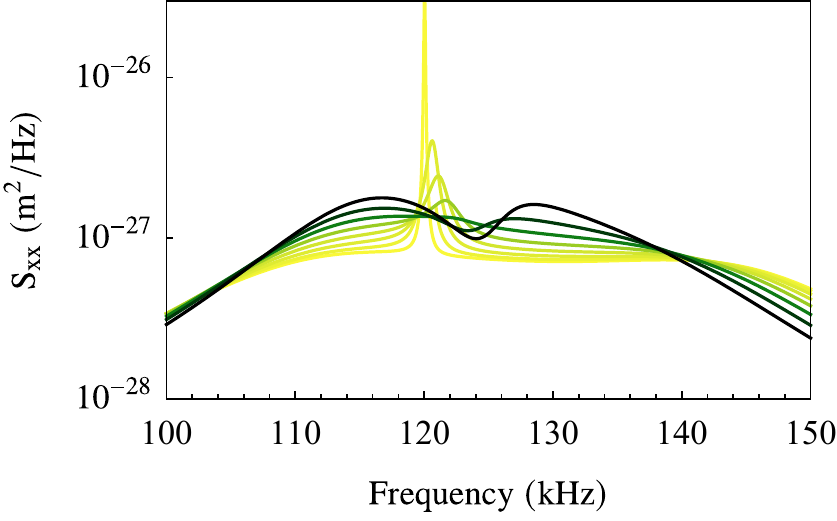}
    \caption{Spectrum of the Stokes sideband, calculated from Eq. (\ref{eq:psd_GBM}) with the following parameters: $\kappa/2 \pi = 57\ \mathrm{kHz}$, $\OmegaX/2 \pi = 131 \ \mathrm{kHz}$, $\OmegaY/2 \pi = 120 \ \mathrm{kHz}$, $g_{\mathrm{max}}/2 \pi = 22.4 \ \mathrm{kHz}$, $\GX/ 2\pi = 5.8 \ \mathrm{kHz}$, $\GY/ 2 \pi = 5.6\ \mathrm{kHz}$, and a polarization angle varying from $85^{\circ}$ (light yellow) to $50^{\circ}$ (dark green) in steps of $5^{\circ}$. The peak of the dark mode evolves into a hole for strong two-dimensional cooling.
    }
    \label{Fig_simulazione}
\end{figure*} 

%\bibliographystyle{apsrev4-2} 
%\bibliography{database}

\end{document}